# VERITAS Observations of Supernova Remnants and Pulsar Wind Nebulae in the Fermi Era

T. B. Humensky, for the VERITAS Collaboration

For full author list, see http://veritas.sao.arizona.edu/conferences/authors?icrc2009

*University of Chicago, Chicago, IL 60637, USA*

Supernova remnants (SNRs) are among the strongest candidates to explain the flux of cosmic rays below the knee around $10^{15}$ eV. Pulsar wind nebulae (PWNe), synchrotron nebulae powered by the spin-down of energetic young pulsars, comprise one of the most populous VHE gamma-ray source classes. Gamma-ray studies in the GeV and TeV bands probe the nature (ions vs. electrons), production, and diffusion of high-energy particles in SNRs and PWNe. For sources that are visible across both the GeV and TeV bands, such as IC 443, the spatial and spectral distribution of gamma rays can be studied over an unprecedented energy range. This presentation will review recent VERITAS results, including studies of Cassiopeia A, IC 443, PSR J1930+1852, and the SNR G106.3+2.7/Boomerang region, and discuss prospects for complementary studies of SNRs and PWNe in the Fermi and VHE gamma-ray bands.

## 1. INTRODUCTION

The last five years have seen a rapid increase in the population of galactic very high energy (VHE, E > 100 GeV) gamma-ray sources. Amongst those with firm identifications at other wavelengths, pulsar wind nebulae (PWNe) are the most common. PWNe form from particles accelerated at the termination shock where a pulsar's wind encounters either the internal medium of a supernova remnant (SNR), or the interstellar medium in the case of older pulsars. They are generally believed to consist of electrons and positrons with energies up to tens or hundreds of TeV diffusing in a magnetized environment. However, neither the nature of the pulsar wind nor the process by which particles are accelerated at the termination shock are very well understood. Similarly, open questions remain surrounding the internal structure of PWNe: in many cases the strength of the magnetic field is not well determined, and how the energetic particles diffuse and cool is poorly known. VHE gamma-ray measurements can shed light on these questions by, for example, constraining the magnetic field when combined with X-ray measurements, or helping to locate cooling breaks in the particle spectrum [1].

Supernova remnants are widely considered to be strong candidates for the sources of galactic cosmic rays, and measurements of nonthermal X-ray spectra from several young SNRs demonstrate that they do accelerate electrons in their shells. Do they accelerate cosmic-ray nuclei as well, and if so how efficient is that acceleration? What maximum energy can be reached? Is diffusive shock acceleration the mechanism, and how important are nonlinear effects in modifying the shock structure and acceleration efficiency? As with PWNe, these questions can be addressed with VHE gamma rays by studying the emission from both young shell-type SNRs and middle-aged SNRs interacting with nearby molecular clouds [1].

VERITAS, the Very Energetic Radiation Imaging Telescope Array System, is an array of four 12-m-diameter imaging air Cherenkov telescopes located at the Fred Lawrence Whipple Observatory (1268 m a.s.l.) on Mt. Hopkins in southern Arizona, USA. Each telescope is equipped with a 3.5° field-of-view camera made of 499 photomultiplier tubes (PMTs). The PMT signals are read out by custom-built 500-MSps flash-ADCs and stored to disk for analysis. A more detailed description of VERITAS is available in [2] and [3]. VERITAS has been fully operational since the fall of 2007, with an effective energy range covering 100 GeV – 30 TeV, an angular resolution of ~0.1° above 1 TeV, and an energy resolution of ~15-20%. The results presented in these proceedings derive from data taken in the first two years of operations, when VERITAS was able to detect a point-like source at the 5σ level with a flux of 1% that of the Crab Nebula in less than 50 hours. Following the relocation of one telescope during the summer of 2009 to optimize the baselines between telescopes, VERITAS can now detect a 1% Crab Nebula flux in less than 30 hours. Details on the VERITAS performance with this upgrade are covered in [4] in these proceedings.

## 2. OBSERVATIONS

As of this Symposium, the VERITAS catalog includes 25 sources in six classes, 10 of which lie in the galactic plane. Of these, four are new detections reported at this meeting, including three discoveries in the VHE band: the blazar RBS 0413; VER J0521+211, most likely associated with the active galaxy RGB J0521.8+2112; the unidentified source TEV J2032+4130 (discovered by HEGRA [5]); and VER J2019+407, most likely associated with the γCygni supernova remnant. The latter two sources were detected during the VERITAS survey of the Cygnus region of the galactic plane, and are discussed in more detail in [6] in these proceedings.

In this paper, we report on four objects that were observed by VERITAS as part of the supernova remnant / pulsar wind nebula Key Science Project, designed to explore the questions discussed earlier. The results include

- Discovery of VHE emission from the pulsar wind nebula associated with PSR J1930+1852 (SNR G54.1+0.3).
- Detection of extended emission associated with SNR G106.3+2.7, which contains the energetic pulsar PSR J2229+6114 and the Boomerang PWN.

- Discovery of extended VHE emission from the SNR IC 443.
- Measurement of the VHE spectrum of the young SNR Cassiopeia A.

## 2.1. PSR J1930+1852 / SNR G54.1+0.3

PSR J1930+1852 has an X-ray morphology [7] that is similar to the Crab Nebula, showing a jet and torus structure. PSR J1930+1852 is young (~2900 yrs) and has a high spin-down luminosity of $1.2 \times 10^{37}$ erg/s. It is ~6.2 kpc distant. Following a hint of a signal in data taken under moonlight in 2007, VERITAS observed PSR J1930+1852 for 31 hours in 2008-09, yielding a $7\sigma$ detection of a point-like source. No extension was observed, indicating a compact VHE nebula, again similar to the Crab Nebula. The VHE emission location agrees well with the pulsar location. The photon spectrum follows a power-law with an index of $2.40 \pm 0.24_{stat} \pm 0.30_{sys}$ over the energy range 250 GeV – 4 TeV, and the integral flux above 1 TeV is 2.5% that of the Crab Nebula. The gamma-ray luminosity in this energy range is $1.8 \times 10^{34}$ erg/s, corresponding to 0.15% of the spin-down luminosity, an efficiency for gamma-ray production that is comparable to other young VHE PWNe such as G0.9+0.1 [8] and Kes 75 [9].

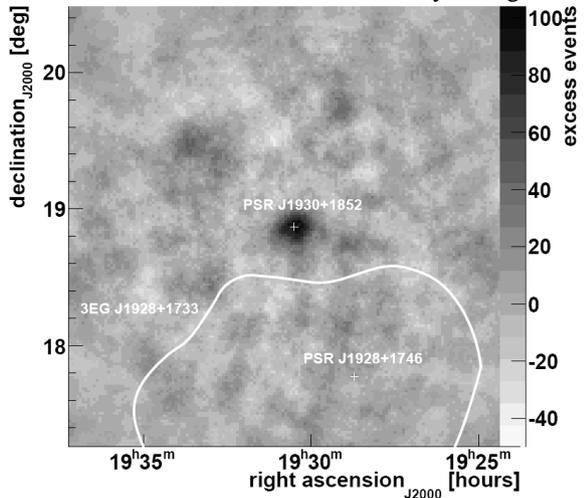
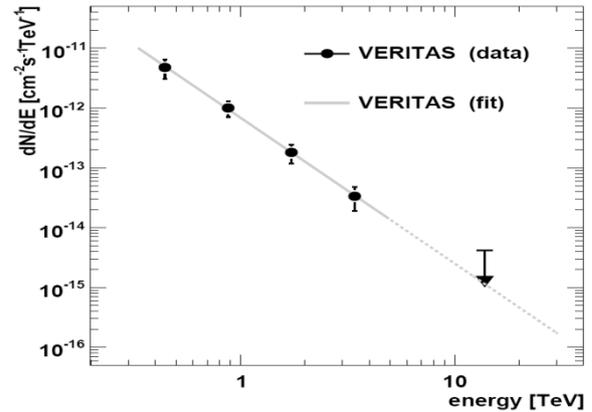

Figure 1: (left) Excess map for the region around PSR J1930+1852 (white cross). The locations of the nearby PSR J1928+1746 and the EGRET unidentified source 3EG J1928+1733 are also indicated. (right) Photon spectrum.

## 2.2. SNR G106.3+2.7 / PSR J2229+6114

The energetic pulsar PSR J2229+6114 was discovered within the error box of 3EG J2227+6122 and has an age of ~10 kyr and a spin-down luminosity of $2.2 \times 10^{37}$ erg/s [10]. Kothes et al. discuss the likely association between this pulsar and SNR G106.3+2.7, and place the distance at 800 pc [11]. Pulsed gamma-ray emission was reported by Fermi [12], and MILAGRO reported the first detection of TeV gamma rays from this region, reporting an extended source at a characteristic energy of 35 TeV [13].

VERITAS observed the SNR G106.3+2.7 field in 2008 for 33 hours live time and found extended emission in the TeV regime with a significance of 7.3 σ (6.0 σ post-trials, for a search within the boundaries of the radio SNR). Figure 2 (left) shows the excess map, indicating the significant extension of the source. The emission profile is somewhat irregular, and spans a region approximately 0.4° × 0.6°. Figure 2 (right) shows the photon spectrum over the energy range 1 – 25 TeV. It is well described by a power law with an index of $2.29 \pm 0.33_{stat} \pm 0.30_{sys}$ and a flux of 5% that of the Crab Nebula above 1 TeV. The extrapolation of this fit up to 35 TeV agrees well with the MILAGRO measurement.

The TeV emission peaks 0.4° from the pulsar location, corresponding to 6 pc at a distance of 800 pc. This is certainly consistent with offset TeV emission observed in other systems, but perhaps inconsistent with the evolutionary scenario presented in [11], in which at the time of the explosion, the progenitor was located near the northeast end of the current radio SNR. In that picture, the SNR encountered a fairly dense medium to the north and east but broke out into a less-dense region to the southwest, allowing rapid expansion in that direction.

An alternative explanation for the TeV emission is provided by the presence of a molecular cloud overlapping the line of site to the SNR: it is observed in CO at a velocity appropriate to the distance to SNR G106.3+2.7, and correlates well with the TeV emission. This spatial overlap suggests that perhaps hadronic cosmic rays have been accelerated in the expanding shell of SNR G106.3+2.7, and are now interacting with the molecular cloud [14].

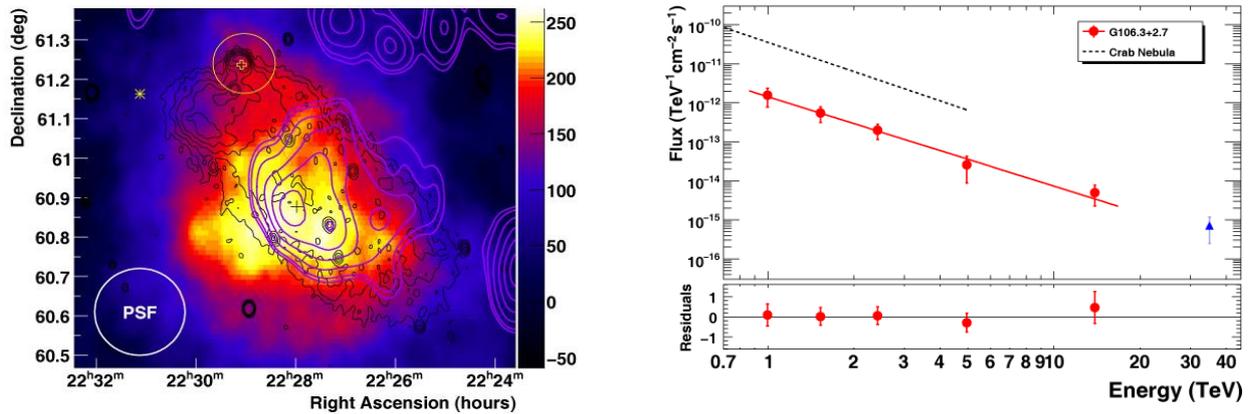

Figure 2: (left) Excess map for the region around SNR G106.3+2.7 / PSR J2229+6114. The black contours indicate the radio shell of the SNR, and the purple contours indicate the density of CO emission. The yellow circle is the Fermi error box, and the open yellow cross indicates the position of the pulsar. The yellow star is the AGILE source 1AGL J2231+6109. (right) Photon spectrum for SNR G106.3+2.7.

## 2.2. IC 443

IC 443 is one of the classic examples of an SNR interacting with a molecular cloud. The cloud appears to form a ring around the SNR, with a portion of it crossing our line of site to IC 443 along an axis from northwest to southeast; the optical emission is suppressed in this region due to absorption [15]. A pulsar wind nebula has been located near the southern edge of the shell [16], though no pulsed emission has yet been found from the neutron star, and the identification of the neutron star as IC 443's progenitor is not yet certain [17,18]. EGRET [19], AGILE [20], and Fermi [12] have reported gamma-ray detections of IC 443, and Fermi presents an updated analysis in these proceedings [21]. In the TeV band, VERITAS and MAGIC both reported emission from IC 443 in 2007 [22,23]. With additional observations, VERITAS reported [24] an extended source centered at (RA, Dec) $6^h16^m51^s + 22°30'11''$ (J2000) $\pm$ 0.03°$_{stat}$ $\pm$ 0.08°$_{sys}$, with an intrinsic extension of 0.16° $\pm$ 0.03°$_{stat}$ $\pm$ 0.04°$_{sys}$. The VHE spectrum is well fit by a power law with a photon index of 2.99 $\pm$ 0.38$_{stat}$ $\pm$ 0.3$_{sys}$ and an integral flux above 300 GeV of (4.63 $\pm$ 0.90$_{stat}$ $\pm$ 0.93$_{sys}$) $\times$ $10^{-12}$ cm$^{-2}$ s$^{-1}$, corresponding to 3.2% of the Crab Nebula flux. The extension is consistent with that seen by Fermi, though the centroids are displaced from one another at ~95% confidence level, with the Fermi emission centered on the remnant and the VERITAS and MAGIC emission overlapping the dense regions of the molecular cloud.

While it may be possible that at least a fraction of the TeV emission arises from relic electrons left behind by the pulsar wind nebula as it moved south through SNR interior, the displacement between the GeV and TeV emission makes it difficult to explain all of the gamma-ray emission via the PWN. The interaction between the SNR shock and the molecular cloud – seen clearly in maser emission from a compact region within the cloud [25] – suggests that a more natural scenario may be that hadronic cosmic rays accelerated within the shock have propagated into the molecular cloud [26].

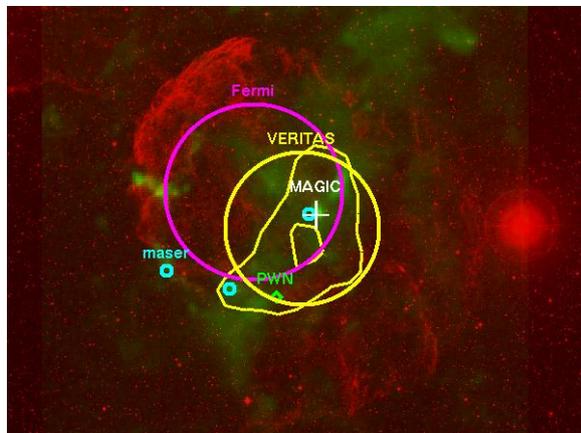

Figure 3: (left) Excess map for the region around IC 443. The red is optical [27] and green is CO [28]. The VERITAS and Fermi one-sigma extensions are indicated by yellow and purple circles, respectively. VERITAS significance contours at 4 and 6 σ are indicated by yellow lines. The MAGIC centroid is indicated by a white cross [23]. Locations of maser emission are indicated by light blue open circles (J. Hewitt, personal communication, 2009), and the pulsar wind nebula is indicated by a green diamond [16]. (right) Photon spectrum for IC 443.

## 2.3. Cassiopeia A

With an age of ~330 years, Cassiopeia A (Cas A) is the youngest of the historical SNRs. Cas A is at a distance of 3.4 kpc and has a diameter of ~5 arcminutes, comparable to the angular resolution of telescopes in the TeV regime. Very-high-energy gamma-ray emission from Cas A was discovered by HEGRA [29] and confirmed by MAGIC [30], and in these proceedings Fermi reports the first detection of Cas A in the GeV energy range [31]. VERITAS observed Cas A for 22 hours live time in 2007, resulting in a detection at the $8.3\sigma$ level. The emission is consistent with a point source, and no constraint can yet be placed on the sites of particle acceleration within Cas A. The spectrum, shown in Figure 4, is a power law over the range 400 GeV – 5 TeV with an index of $2.61 \pm 0.24_{stat} \pm 0.20_{sys}$ and an integral flux above 1 TeV of 3.5% that of the Crab Nebula. The question of whether the particles producing the gamma rays are electrons or hadrons remains open, but the broad-band gamma-ray spectrum, combining Fermi and VERITAS, will be very constraining in the coming years.

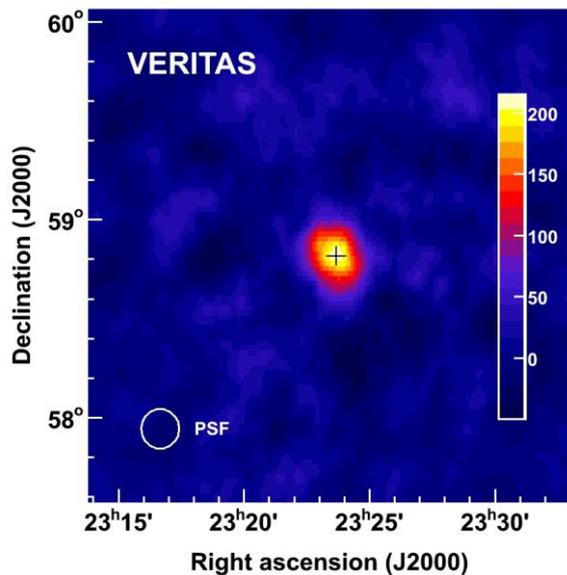
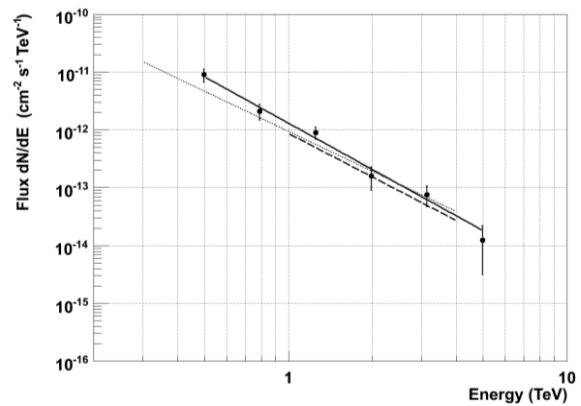

Figure 4: (left) Excess map for the field surrounding Cassiopeia A. (right) Cassiopeia a photon spectrum. The data points are from VERITAS. The solid line represents the best-fit power law spectrum to the VERITAS points, while the dashed line is from HEGRA and the dotted line is from MAGIC.

## 3. SUMMARY AND OUTLOOK

Observations of northern SNRs and PWNe with VERITAS have proven extremely fruitful over the last several years, with discoveries of TeV emission from PSR J1930+1852 and (with MAGIC) IC 443, and strong detections of SNR G106.3+2.7 and Cassiopeia A. Looking forward, the strong detections of IC 443 and Cas A by Fermi show great promise for broad-band gamma-ray studies of these sources. Future deep observations of all of these sources with VERITAS will enhance photon statistics, allowing refined measurements of spectra and improved morphological studies of the extended sources IC 443 and SNR G106.3+2.7.

## Acknowledgments

This research is supported by grants from the U.S. National Science Foundation, the U.S. Department of Energy, and the Smithsonian Institution, by NSERC in Canada, by Science Foundation Ireland, and by STFC in the U.K. The excellent work of the technical support staff at the FLWO and the collaborating institutions is acknowledged.